\documentclass[%
 reprint, 
 amsmath,amssymb,
 aps, physrev,
]{revtex4-2}

\usepackage{graphicx}
\usepackage{dcolumn}
\usepackage{bm}
\usepackage{threeparttable}
\usepackage{float}

\usepackage{xcolor}
\definecolor{myorange}{RGB}{220,80,20}

\begin{document}

\preprint{APS/123-QED}

\title{ {A Microphysically Inspired Approach to Dark Matter--Dark Energy Interactions:
First bounds on dark-sector scattering cross sections}}

\author{A.A. Escobal$^{1,2}$}
 \email{Contact author: anderson.aescobal@outlook.com}
 \author{F.B. Abdalla$^{1,2}$}%
 \email{Contact author: filipe.abdalla@gmail.com}
 \author{J.F. Jesus$^{3,4}$}
 \author{E. Abdalla$^{5,7,8}$}
 \author{C. Feng$^{1,2}$}
\author{J. A. S. Lima$^6$}
\author{O. P. F. Piedra$^{7}$}
\affiliation{$^{1}$
 Department of Astronomy, University of Science and Technology of China, Hefei, Anhui 230026, China.\\
 $^2$ School of Astronomy and Space Science, University of Science and Technology of China, Hefei, Anhui 230026, China. \\
 $^3$Department of Science and Technology, Universidade Estadual Paulista (UNESP), Instituto de Ciências e Engenharia,  Itapeva, SP, 18409-010, Brazil.\\
 $^4$Department of Physics, Universidade Estadual Paulista (UNESP), Faculdade de Engenharia e Ci\^encias de Guaratinguet\'a, Guaratinguet\'a, SP, 12516-410,  Brazil.\\
$^5$Department of Physics, Universidade de S\~ao Paulo, 05508-900, S\~ao Paulo, SP, Brazil.\\
$^6$Department of Astronomy, Universidade de S\~ao Paulo, 05508-900, S\~ao Paulo, SP, Brazil. \\
$^7$Department of Physics, Center for Exact and Natural Sciences, Federal University of Paraíba, 58059-970, João Pessoa, Brazil\\
$^8$Paraíba State University, 351 Baraúnas Street, University District, Campina Grande, Brazil}

\date{\today}

\begin{abstract}
The observational tension regarding the value of the Hubble constant ($H_0$) has motivated the exploration of alternative cosmological scenarios, including Interacting Dark Energy models. However, the majority of IDE models studied in the literature rely on phenomenological interaction terms proportional to the Hubble parameter (e.g., $Q\propto H\rho$), which lack a clear microphysical justification and often suffer from large-scale instabilities. In this work, we propose and investigate a "bottom-up" IDE model where the interaction is  {motivated by} particle physics collision processes, taking the  {effective} form $Q\propto\rho^2$. This interaction  {effectively represents} a reversible annihilation/creation process between Dark Matter and Dark Energy, motivated by the Boltzmann equation. We test this model against a combination of background cosmological data, including Type Ia Supernovae (Pantheon Plus), Cosmic Chronometers, Baryon Acoustic Oscillations (DESI DR2), and the CMB acoustic-scale prior ($\ell_A$) from Planck18. We find that the model is consistent with the data, yielding a Hubble constant of $H_0=68.9^{+1.5}_{-2.0}$ km s$^{-1}$ Mpc$^{-1}$ for the combined analysis. The dimensionless interaction rate coefficients are constrained to be small, with upper limits of $A < 8.04\times10^{-21}$ ( {for the forward dark-sector channel}) and $B < 4.28\times10^{-3}$ ( {for the effective inverse channel}) at 95\% confidence level. Since the interaction model is parameterized by the expansion rate, these bounds on $H_0$, $A$, and $B$  {can be mapped} into a strict limit on the thermally-averaged annihilation cross-section per unit of mass. The constraints on the coupling $A$ imply that, if such a collisional interaction exists, the effective dark-matter annihilation cross section per unit mass is highly suppressed relative to the cosmological expansion rate. In contrast, the corresponding  {effective} dark-energy contribution, governed by $B$, is constrained at the sub-percent level in dimensionless units.
\end{abstract}

\maketitle

\textit{Introduction} --- Currently, the best description of observed cosmology is given by the flat $\Lambda$ Cold Dark Matter model ($\Lambda$CDM-flat). While it successfully describes most astronomical observations, both from the early Universe, such as primordial nucleosynthesis data \cite{ParticleDataGroup:2022pth} and the Cosmic Microwave Background (CMB) \cite{Planck18}, and from the late Universe, such as measurements of Type Ia supernovae (SNe Ia) \cite{Brout:2022vxf} and Hubble parameter data, $H(z)$ \cite{Moresco:2022phi}, it faces several critical challenges \cite{Abdalla:2022yfr}. The most pressing of these is the so-called $H_0$ tension \cite{Perivolaropoulos:2021jda, Brito:2024bhh}, a statistically significant discrepancy around ($\sim 4-6\sigma$) between the value of the Hubble constant measured from early Universe probes and that measured by local late Universe observation probes.

The current statistical tensions, along with other issues within the standard model, present an opportunity to explore alternative models that may arise from modifications in gravitational theory or changes in the matter-energy content composing the dark sector. By maintaining General Relativity and altering the material content, we consider models with interaction between dark energy and dark matter, known as interacting dark energy (IDE) models. The possibility of interaction between dark energy and dark matter is an intriguing avenue for studying the dark sector of the Universe as well as implications within particle physics. IDE cosmological models have been extensively discussed in the literature \cite{Brito:2024bhh,Lima2004,Gavela:2010tm,Farrar:2003uw,vonMarttens:2018bvz,Salvatelli:2014zta,Shafieloo:2016bpk,Marcondes:2016reb,Wang:2016lxa,vomMarttens:2016tdr,Santos:2017bqm,Yang:2017zjs,Cid:2018ugy,Landim:2019lvl,DiValentino:2019ffd,Jesus:2020tby,Lucca:2020zjb,vonMarttens:2020apn,Anchordoqui:2021gji,Gariazzo:2021qtg,Nunes:2022bhn,Bernui:2023byc,Zhai:2023yny,Hoerning:2023hks,Wang:2024vmw}. Considering possible couplings between these unknown components, energy transfer may occur from dark energy to dark matter or vice versa. More generally, both processes may operate simultaneously, leading to a bidirectional exchange within the dark sector.

IDE models are natural candidates for addressing the Coincidence Problem, as the ratio of the dark sector components is dynamic. Recently, IDE models have been extensively applied to address the $H_0$ problem \cite{Brito:2024bhh,DiValentino:2019ffd,Anchordoqui:2021gji,Nunes:2022bhn,Bernui:2023byc}. However, the vast majority of IDE models explored in the literature are ``top-down'' phenomenological \emph{ansatzes} \cite{Wang:2016lxa,Yang:2017zjs,DiValentino:2019ffd,Jesus:2020tby,Lucca:2020zjb,Nunes:2022bhn,Bernui:2023byc,Zhai:2023yny,Hoerning:2023hks}. The most common forms, such as terms proportional to the Hubble function times the energy density of some material component (e.g., $Q \propto H{\rho}$), are motivated by dimensional analysis rather than a fundamental microphysical theory. This popular approach faces significant theoretical challenges: it relies on a ``non-local'' interaction dependent on the global expansion rate ($H$) and, as demonstrated in \cite{Valiviita:2008iv,He:2008si}, can suffer from catastrophic large-scale instabilities in the early universe. While some authors \cite{vonMarttens:2018iav,Figueruelo:2026eis} have considered a non-linear interaction term $Q \propto H\rho^2$, no justification based on first principles has been presented.

In contrast, this work explores a ``bottom-up'' model. Based on the particle physics framework of equilibrium processes \cite{Kolb:1990vq}, we propose and investigate an interaction term $Q$  {motivated by} the microphysics of local particle collisions.
This term represents a reversible collision process at the field level,  {schematically} $\chi + \chi \longleftrightarrow \phi + \phi$, where $\chi$ represents the dark matter field and $\phi$ the dark energy field, corresponding to local, number changing interactions within the dark sector.
As an effective description grounded in microphysics, this formulation is  {more physically motivated} than the common interaction terms of the form $Q \propto H\rho$.

The primary goal of this paper is to test this new, microphysically-motivated IDE model against cosmological observations. We explore its capacity to  {impact the inferred value of $H_0$} by performing a statistical analysis to obtain observational constraints on its parameters, particularly the interaction coefficients.
For this, we use a combined dataset of Cosmic Chronometers (CCs) \cite{Moresco:2022phi}, Type Ia Supernovae (SNe Ia) from Pantheon Plus \cite{Brout:2022vxf}, more recent measurements of Baryon Acoustic Oscillations (BAO) from DESI DR2 \cite{DESI:2025zgx}, and the CMB acoustic-scale prior ($\ell_A$) \cite{Chen:2018dbv}.

\textit{\label{DFIDE}Dynamic Formulation of the IDE Model} --- To evaluate the interaction between dark matter and dark energy, we postulate an interaction term $Q_i^\nu$.
The non-conservation of the energy-momentum tensor for these individual components is thus
\begin{align}
\nabla_\mu T_{{i}}^{\mu \nu} &=Q^\nu_{i} \,.
\end{align}
Considering the interaction 4-vector $Q^\nu_i = [Q_i,0,0,0]^T$, as implemented in \cite{Wang:2016lxa}, and defining $Q_{\text{DM}} = Q = -Q_{\text{DE}}$
within the context of the FLRW metric, the background-level conservation laws for the dark matter ($\bar{\rho}_{\text{DM}}$) and dark energy ($\bar{\rho}_{\text{DE}}$) densities are:
\begin{align}
\dot{\bar{\rho}}_{\text{DM}} + 3H\bar{\rho}_{\text{DM}} &={Q}\,,\label{eqDM}\\
\dot{\bar{\rho}}_{\text{DE}}+3H{\bar{\rho}}_{\text{DE}}(1+w) &=-{Q}\,.\label{eqDE}
\end{align}

Here, the overdot denotes the derivative with respect to cosmic time, $w$ is the dark energy equation of state parameter, and $H$ is the Hubble parameter. We have $Q>0$ when dark energy is changing into dark matter, while $Q<0$ corresponds to DM changing into DE. Baryons and radiation are assumed to be conserved separately and evolve as in the standard $\Lambda$CDM model.

To explore new routes in the dark sector, we analyze an interaction term that differs from those commonly adopted in the literature. As discussed in the Introduction, the most widely-used forms (e.g., $Q \propto H\bar{\rho}$) are largely phenomenological \emph{ansatzes}, motivated by dimensional analysis rather than by a first-principles microphysical theory, and their dependence on the global expansion rate $H$ can render them prone to large-scale instabilities \cite{Valiviita:2008iv,He:2008si}. Instead, we motivate the quadratic form $Q \propto \bar\rho^2$ by analogy with the standard Boltzmann collision term for particle annihilation \cite[Chapter 5]{Kolb:1990vq}, which depends only on local fluid properties. For the annihilation of a generic particle $\psi$, the evolution of its number density $n_\psi$ is given by \cite[Eq.~5.24]{Kolb:1990vq}:
\begin{align}
    \label{eq:boltzmann_n}
    \frac{d n_\psi}{dt} + 3Hn_\psi = - \langle\sigma|v|\rangle \left[ n_\psi^2 - (n_\psi^{EQ})^2 \right],
\end{align}
where $\langle\sigma|v|\rangle$ is the thermally-averaged annihilation cross-section, and the terms $-\langle\sigma|v|\rangle n_\psi^2$ and $+\langle\sigma|v|\rangle (n_\psi^{EQ})^2$ represent the annihilation and creation processes, respectively.

In our case the relevant reaction is the reversible $2\!\rightarrow\!2$ process $\chi+\chi\leftrightarrow\phi+\phi$, which couples the two dark species. The pair of Boltzmann equations for their number densities reads
\begin{align}
\dot n_{\chi} + 3Hn_{\chi} &= -\langle\sigma|v|\rangle_{A}\,n_{\chi}^2 + \langle\sigma|v|\rangle_{B}\,n_{\phi}^2 \,,\\
\dot n_{\phi} + 3Hn_{\phi} &= +\langle\sigma|v|\rangle_{A}\,n_{\chi}^2 - \langle\sigma|v|\rangle_{B}\,n_{\phi}^2 \,,
\end{align}
so that the role played by the equilibrium term $(n_\psi^{EQ})^2$ in Eq.~\eqref{eq:boltzmann_n} is taken here by the density of the \emph{partner} species: the creation of dark matter is sourced by $\phi\phi$ annihilation, and conversely. Converting to mass densities via $\bar\rho = m\,n$  {for non-relativistic particle-like components} and collecting the energy transfer into a single term $Q$ entering Eqs.~\eqref{eqDM}--\eqref{eqDE} with opposite signs, we obtain 
\begin{align}\label{Q1}
    Q = -A\frac{H_0}{\bar\rho_{\text{c},0}}\bar{\rho}_{\text{DM}}^2+B\frac{H_0}{\bar\rho_{\text{c},0}}\bar{\rho}_{\text{DE}}^2 \,.
\end{align}
Here, $A$ and $B$ are the dimensionless and positive  {effective interaction rate coefficients}, and $\bar\rho_{c,0}$ is the critical energy density at the present epoch.
The term $Q$ models the dark-sector interaction as a reversible collision process and represents the net balance of the reaction; its sign, and thus the net direction of energy flow, depends on the dynamical competition between the forward and inverse channels throughout cosmic history. The microphysical foundations of Eq.~\eqref{Q1} --- the renormalizable Lagrangian that generates the $\chi\chi\leftrightarrow\phi\phi$ process, the explicit conditions under which the quadratic form emerges, the resulting thermally-averaged cross section, and its connection to the coefficients $A$ and $B$ --- are presented in Appendix~\ref{app:micro}.

\textit{Effective Field-Theory Interpretation of the Dark-Sector Interaction \label{microphysics}} --- At the microscopic level, the interaction considered in this work can be interpreted within an effective field theory framework involving two dark-sector fields, denoted by $\chi$ and $\phi$, associated with dark matter and dark energy, respectively.

For bosonic dark matter, a minimal and renormalizable realization of the interaction is provided by quartic operators in the Lagrangian density. In the case of scalar dark matter \cite{Escobal:2020rfn}, this takes the form $\mathcal{L}_{\mathrm{int}} \propto \chi^2\phi^2$, while for vector dark matter one may write $\mathcal{L}_{\mathrm{int}} \propto (\chi_\mu\chi^\mu)\phi^2$. Both interactions are Lorentz invariant, local, and renormalizable  {at the effective level}, and mediate reversible $2\!\rightarrow\!2$ processes of the type $\chi+\chi\leftrightarrow\phi+\phi$.
For Majorana fermionic dark matter, an analogous  {effective} interaction can arise from $\mathcal{L}_{\mathrm{int}} \sim (\bar{\chi}\chi)\phi^2/\Lambda_c$, where $\Lambda_c$ is an energy cutoff scale, possibly related to the dark-sector particle. This operator leads to the same quadratic dependence of the interaction term at the coarse--grained, background level.
 Axion or axion--like dark matter \cite{Liu:2025qwf} may also be accommodated within this framework, with the important caveat that, denoting the axion field by $\theta$, couplings such as $\theta^2\phi^2$ are forbidden by the axion-like shift symmetry; an interaction can then arise only through breaking of this symmetry.

Although written as contact quartics, these operators need not be fundamental. As shown explicitly in Appendix~\ref{app:micro}, integrating out a heavy scalar mediator $S$ of mass $M$ that couples to the dark bilinears $\chi^2$ and $\phi^2$ (but not to $\chi\phi$) generates the cross operator $\chi^2\phi^2$ with effective coupling $\lambda_{\rm eff}=g_\chi g_\phi/M^2$, while \emph{simultaneously} producing the $\chi^4$ and $\phi^4$ self-couplings  {that can help maintain kinetic equilibrium when their rates are sufficiently large}.
In this completion the conversion cross section scales as $\langle\sigma|v|\rangle\propto M^{-4}$, so that the coefficients obey $A,B\propto M^{-4}$.  {Provided the couplings $g_\chi,g_\phi$ and the masses are held fixed and the contact regime $M\gg2m_\chi$ applies, the observational smallness of $A$ and $B$ reported below can be read as the signature of a \emph{heavy} mediator---in direct analogy with the Fermi suppression $G_F\sim g^2/M_W^2$ of the weak interaction---rather than as a fine-tuned input. More generally, since $A$ constrains the combination $(g_\chi g_\phi)^2/(M^4 m_\chi^3)$, a small coupling could equally reflect weak couplings, a large $m_\chi$, or kinematic threshold suppression; the heavy-mediator reading is therefore one consistent interpretation rather than a unique inference, and follows from a well-defined Lagrangian.}

For this interaction to be operative, the dark matter field $\chi$ must admit number--changing self-annihilation processes. This requirement naturally selects $\chi$ to be a bosonic field, such as a scalar or vector degree of freedom, or a Majorana fermion. In particular, standard Dirac fermionic dark matter would require additional internal quantum numbers or higher-dimensional operators to allow for $\chi\chi$ annihilation without violating Fermi statistics.

The dark energy component $\phi$ is assumed to be a dynamical field rather than a strict cosmological constant. This assumption is essential, as a vacuum energy term does not support particle-like excitations or number--changing interactions. In this framework, $\phi$ may be interpreted as a light scalar field or condensate, capable of participating in local interactions while maintaining an effective equation of state close to $w \simeq -1$ at the background level. The interaction considered here therefore excludes models in which dark energy is exactly constant or purely geometric in origin.

It is important to emphasize that not all dark-sector interactions are compatible with the present framework. Linear decay channels, such as $\chi \rightarrow \phi$ or $\phi \rightarrow \chi$, are explicitly excluded, as they are not collisional and would generate a qualitatively different $Q\propto\bar\rho$ phenomenology. Likewise, interaction terms proportional to a single power of the energy density do not arise from local $2\!\rightarrow\!2$ processes and are not supported by this effective field theory interpretation. The quadratic dependence of the interaction term thus reflects the underlying assumption that energy exchange in the dark sector is dominated by local, two-particle collisions.

Within these assumptions, Eq.~\eqref{Q1} is the background--level manifestation of an underlying microphysical interaction. The coefficients $A$ and $B$ parametrize the effective efficiency of the forward and inverse processes after averaging over phase space and macroscopic field configurations, and need not coincide despite originating from a symmetric microscopic interaction.  {In particular, whereas $A$ maps onto the forward rate more directly (cold dark matter, $\bar\rho_{\rm DM}=\bar\rho_\chi$), $B$ is genuinely effective: because only the particle-like component of dark energy can scatter [condition (V3)], the inverse channel sources $\bar\rho_{\phi,{\rm part}}^2=f_{\rm part}^2\,\bar\rho_{\rm DE}^2$, with $f_{\rm part}\equiv\bar\rho_{\phi,{\rm part}}/\bar\rho_{\rm DE}$, so $B$ absorbs the factor $f_{\rm part}^2$ together with the inverse-channel threshold. A single transfer term $Q$ with opposite signs in Eqs.~\eqref{eqDM}--\eqref{eqDE} moreover presumes a closed background energy balance, exact only for $m_\chi\simeq m_\phi$; any residual per-event mismatch is likewise absorbed into the effective coefficients.}
 Appendix~\ref{app:micro} presents the renormalizable Lagrangian that generates the process, the conditions under which the quadratic form emerges, and the connection of $A$ and $B$ to the thermally-averaged cross sections.

\textit{\label{data}Methodology and Observational Data Sample} --- The models were analyzed employing the nested sampling method via the \texttt{dynesty} package \cite{dynesty} to sample from our probability distribution within the $n$-dimensional parameter space. Observational data play a crucial role in validating cosmological models and analyzing current cosmological tensions, since the precise constraints imposed by new data samples select the IDE models that best represent the observed Universe.

In this work we combine four background probes spanning the late and early Universe. For the Type Ia Supernovae (PP) we used the Pantheon$+$ data \cite{Brout:2022vxf}, which consists of 1701 light curves of $1550$ distinct SNe Ia in the redshift range $0.001<z<2.26$. As a probe of the expansion history, we also employed Cosmic Chronometers (CCs): in terms of redshift, the Hubble function can be expressed as $H(z)=-\frac{1}{1+z}\frac{dz}{dt}$, and we used the most recent sample of $32$ astrophysical measurements of $H(z)$ with covariance, obtained by estimating the differential ages of galaxies \cite{Moresco:2022phi}, in the redshift range $0.07<z<1.97$. To anchor the early-Universe physics, we incorporate a prior from the CMB (PCMB) provided by the Planck 18 collaboration \cite{Planck18}; as discussed in \cite{Chen:2018dbv}, CMB distance priors provide a good fit for cosmological models without the need to use perturbation theory to study the entire CMB power spectrum, and in our analyses we use the prior on the acoustic scale $\ell_A$ defined therein. Finally, we utilize the most recent sample of Baryon Acoustic Oscillation (BAO) measurements from the DESI collaboration DR2 \cite{DESI:2025zgx}, containing measurements of $D_{M}/r_{s,drag}$, $D_{H}/r_{s,drag}$ and $D_{V}/r_{s,drag}$\footnote{Let $D_{M}$ be the transverse comoving distance, $D_{H}$ the Hubble distance, $D_{V}$ the isotropic BAO distance, and $r_{s,drag}$ the comoving sound horizon at the drag epoch. For the calculation of the drag epoch redshift, $z_d$, we adopt the approximation described in \cite{Hu:1995en}.} covering the redshift range $0.2950 < z < 2.33$.

\textit{\label{Res}Results} --- To assess the viability of the proposed IDE model, we constrain its free parameters using the observational datasets and statistical framework introduced above. We perform a Bayesian analysis, sampling the posterior probability distributions of the free parameters for each combination of observational datasets.

We adopt wide, flat priors for the standard cosmological parameters $(\Omega_{\text{DM0}}, \Omega_{\text{b0}}, w, H_0)$. Regarding the interaction rate coefficients, given the lack of prior knowledge about their expected magnitude, we adopt a non-informative {Jeffreys Prior} (or log-uniform prior). This choice ensures scale invariance, effectively assigning equal prior probability to each order of magnitude. Consequently, we sample the exponents $A$ and $B$ from a uniform distribution, such that the physical interaction rates are given by $10^A$ and $10^B$. The complete set of priors is summarized in Table \ref{tab:Prior}.

\begin{table}[h]
\caption{Priors on the free parameters of the IDE model.}
\begin{tabular}{c c}
\hline\hline
Parameter & Prior Range \\
\hline
$\Omega_{\text{DM0}}$ & $[0, 1.0]$ \\
$\Omega_{\text{b0}}$ & $[0.01, 0.085]$ \\
$w$ & $[-2.5, -0.1]$ \\
$\log_{10}A$ & $[-60, 10]$ \\
$\log_{10}B$ & $[-60, 10]$ \\
$H_0$ [$\mathrm{km\,s^{-1}\,Mpc^{-1}}$] & $[20, 120]$ \\
\hline\hline
\end{tabular}
\label{tab:Prior}
\end{table}

\begin{figure*}
    \centering
    \includegraphics[width=0.45\linewidth]{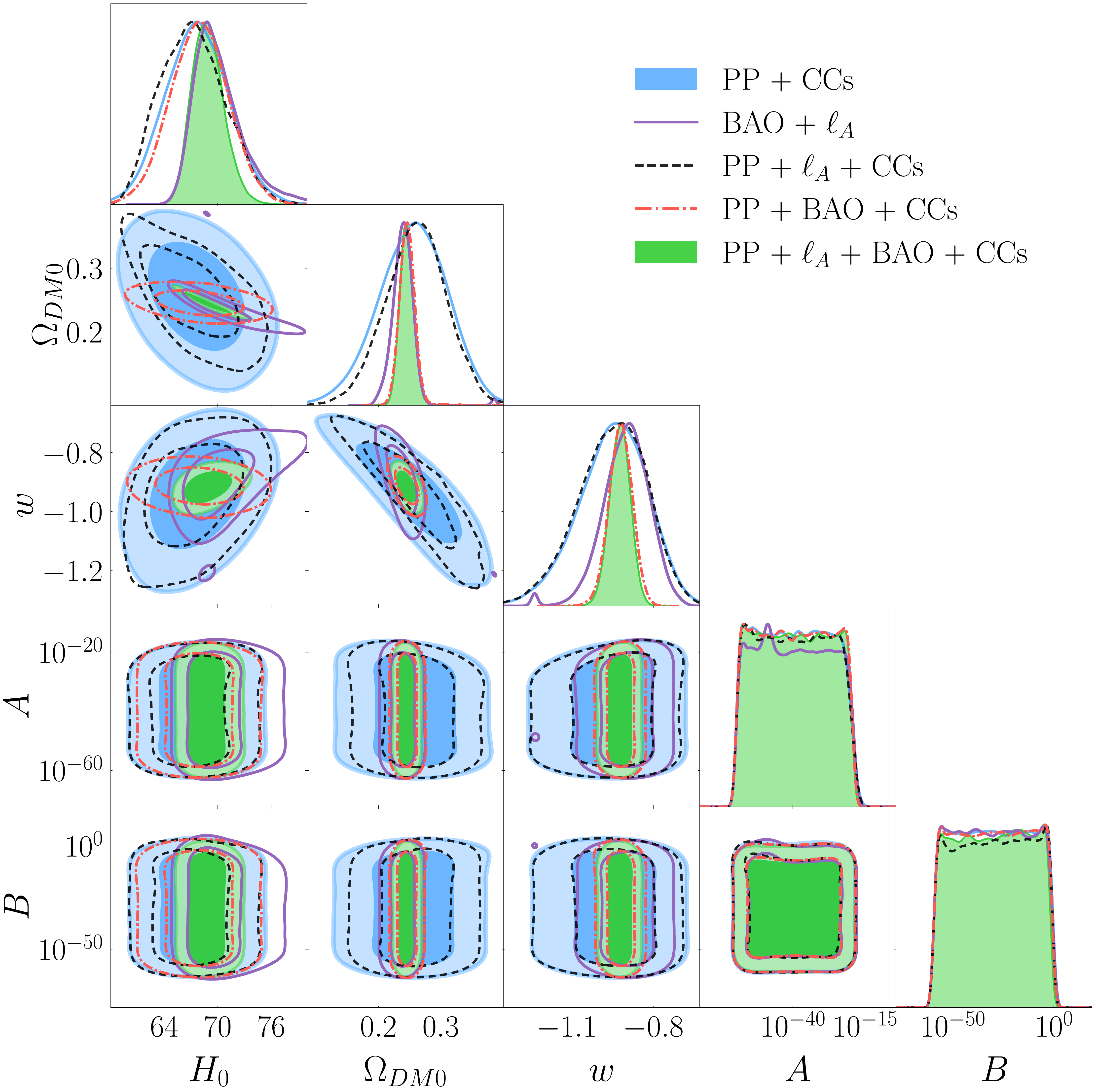}\includegraphics[width=0.45\linewidth]{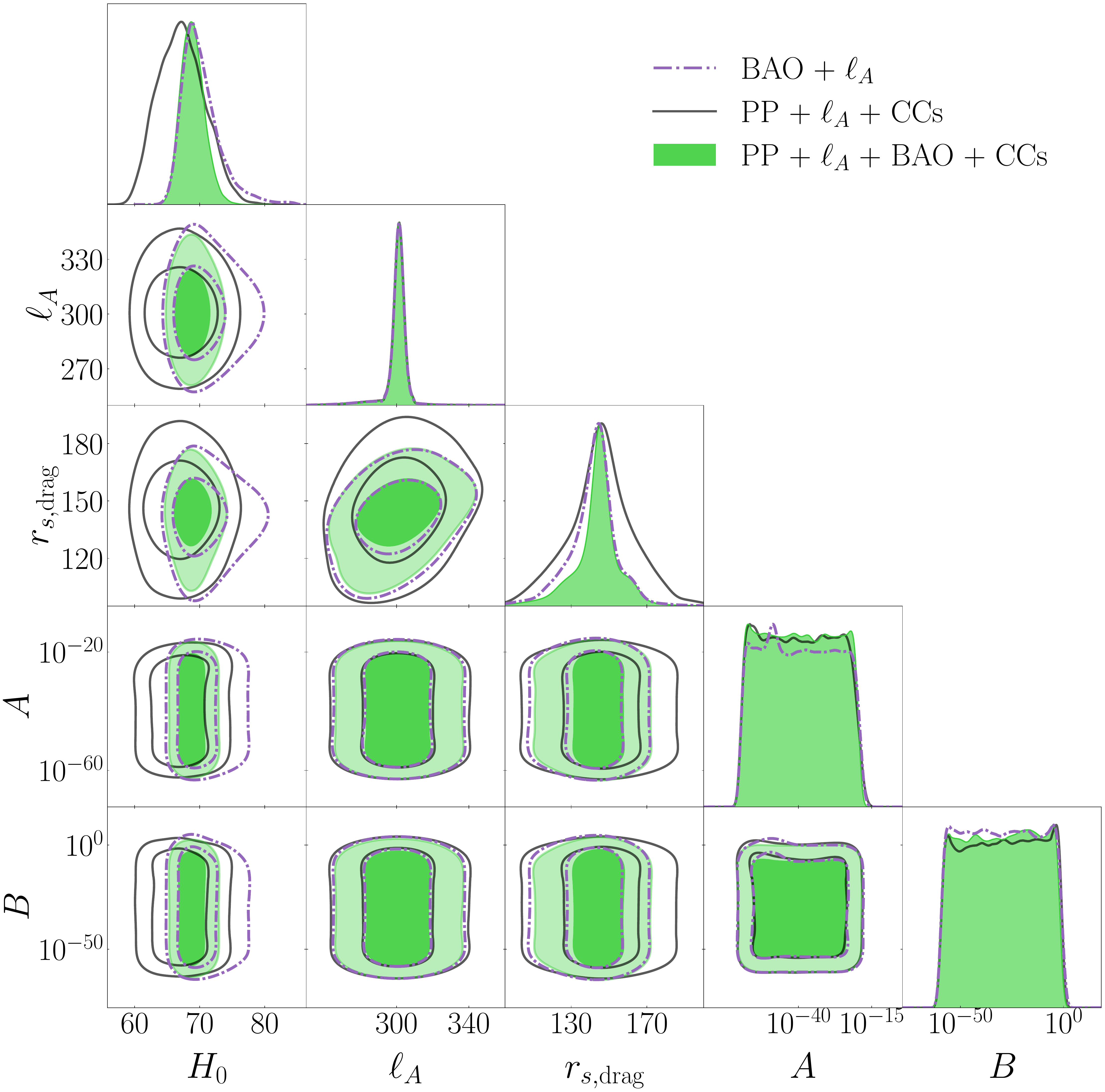}
    \caption{Marginalized 1$\sigma$ and 2$\sigma$ confidence contours and posterior distributions for the free parameters.  {Left:} Constraints on the primary model parameters: $H_0$, $\Omega_{\text{DM0}}$, $w$, and the interaction coefficients $A$, $B$.  {Right:} Constraints highlighting the derived acoustic scales: $H_0$, $\ell_A$, $r_{s,drag}$, alongside the interaction coefficients $A$, $B$. Note that for clarity, not all individual dataset combinations are shown in every panel.}
    \label{fig:contours}
\end{figure*}

The marginalized constraints for the parameters $H_0$, $\Omega_{\text{DM0}}$, $\Omega_{\text{b0}}$, $w$, $\ell_A$, and $r_{s,drag}$ are presented in Table \ref{tab:observational_constraints_final_limits} at 68\% confidence level (CL). For the interaction coefficients $\log_{10}A$ and $\log_{10}B$, the reported values correspond to the upper limits at 95\% CL. We also display the 1$\sigma$ and 2$\sigma$ confidence contours for the full combination of data in Figure \ref{fig:contours}.

\begin{table*}
\centering
\caption{Marginalized mean values and 68\% confidence level constraints on the free parameters and derived quantities. For the interaction coefficients $A$ and $B$, the reported values are upper limits at 95\% confidence level.}
\label{tab:observational_constraints_final_limits}
\footnotesize
\setlength{\tabcolsep}{3.0pt}
\renewcommand{\arraystretch}{1.8}
\begin{tabular}{l|cccccccc}
\hline\hline
 {Dataset} &  {$H_0$} &  {$\Omega_{\text{DM0}}$} &  {$\Omega_{\text{b0}}$} &  {$w$} &  {$A$} &  {$B$} &  {$\ell_A$} &  {$r_{s,drag}$} \\
\hline
PP + CCs & $67.7\pm 3.6$ & $0.256\pm 0.058$ & $0.050\pm 0.023$ & $-0.94^{+0.13}_{-0.10}$ & $< 10^{-20.118}$ & $< 10^{-1.228}$ & --- & --- \\
BAO + $\ell_A$ & $70.2^{+1.6}_{-3.3}$ & $0.240\pm 0.023$ & $0.0572^{+0.0070}_{-0.011}$ & $-0.897^{+0.085}_{-0.070}$ & $< 10^{-19.700}$ & $< 10^{-1.541}$ & $300.87^{+0.82}_{-0.013}$ & $142^{+12}_{-10}$ \\
PP + $\ell_A$ + CCs & $67.2^{+3.2}_{-4.0}$ & $0.260\pm 0.049$ & $0.0467^{+0.015}_{-0.0068}$ & $-0.95^{+0.13}_{-0.10}$ & $< 10^{-20.098}$ & $< 10^{-1.250}$ & $300.87^{+0.85}_{-0.056}$ & --- \\
PP + BAO + CCs & $67.9\pm 3.2$ & $0.246\pm 0.013$ & $0.0522^{+0.0054}_{-0.0049}$ & $-0.914\pm 0.041$ & $< 10^{-19.988}$ & $< 10^{-1.571}$ & --- & $146^{+13}_{-12}$ \\
PP + $\ell_A$ + BAO + CCs & $68.9^{+1.5}_{-2.0}$ & $0.245\pm 0.011$ & $0.0534\pm 0.0046$ & $-0.916\pm 0.036$ & $< 10^{-20.095}$ & $< 10^{-2.369}$ & $300.92^{+0.88}_{-0.22}$ & $144^{+11}_{-8.7}$ \\
\hline\hline
\end{tabular}
\end{table*}

We analyze the constraints obtained from the different combinations of cosmological probes considered in this work. This step is crucial to check for internal consistency among the datasets within the framework of our IDE model.

Looking at the Hubble constant, $H_0$, we observe a broad compatibility among the data combinations, albeit with a mild spread driven by the inclusion of BAO and the acoustic-scale prior. The late-time combinations without BAO prefer lower values: PP + CCs yields $H_0 = 67.7 \pm 3.6$ km s$^{-1}$ Mpc$^{-1}$, and PP + $\ell_A$ + CCs gives $67.2^{+3.2}_{-4.0}$ km s$^{-1}$ Mpc$^{-1}$. In contrast, the combinations that incorporate BAO favor somewhat higher values, with PP + BAO + CCs giving $67.9 \pm 3.2$ km s$^{-1}$ Mpc$^{-1}$ and BAO + $\ell_A$ yielding the highest central value, $70.2^{+1.6}_{-3.3}$ km s$^{-1}$ Mpc$^{-1}$. All values remain mutually consistent within their $1\sigma$ uncertainties, and the interaction model accommodates the data without introducing tension among the probes.

Regarding the interaction sector, we find that the interaction parameters are constrained only as upper limits. The annihilation coefficient $A$ is bounded at a comparable level across all combinations, with $\log_{10}A$ ranging from $< -19.70$ (BAO + $\ell_A$) to $< -20.12$ (PP + CCs). The creation coefficient $B$ is somewhat more sensitive to the dataset: its upper limit ranges from $\log_{10}B < -1.23$ (PP + CCs) to $\log_{10}B < -2.37$ for the full combination, the tightest bounds being obtained whenever BAO is included.

For the full combination (PP + $\ell_A$ + BAO + CCs), we obtain a measurement of the Hubble constant, $H_0 = 68.9^{+1.5}_{-2.0}$ km s$^{-1}$ Mpc$^{-1}$. Regarding the interaction parameters in the joint analysis, we find strict upper limits of $A < 8.04\times10^{-21}$ and $B < 4.28\times10^{-3}$ at 95\% confidence level.

To evaluate the goodness of fit, we calculate the reduced chi-squared $\chi^2_{\nu}$ \footnote{Defined as: $\chi^2_\nu\equiv\frac{\chi^2_{min}}{n-p}$ \cite{Brito:2024bhh}}. For the full combination the IDE model yields $\chi^2_{\nu}= 0.8756$, comparable to $\chi^2_{\nu,\text{F}\Lambda\text{CDM}}= 0.8793$, with both below unity. The interaction thus provides a statistically viable fit,  {with a reduced $\chi^2$ comparable to that of $\Lambda$CDM}. This viability is what makes the limits on $A$ and $B$ in Table~\ref{tab:observational_constraints_final_limits} meaningful bounds on an admissible dark-sector interaction, rather than artifacts of a disfavored model. A breakdown across individual probes is given in Appendix~\ref{API}.

We also investigate the impact of the interaction on the acoustic angular scale, $\ell_A$, and the comoving sound horizon at the drag epoch, $r_{s,drag}$. We observe that the interaction strength is correlated with the sound horizon. The combinations that include the acoustic-scale prior pin $\ell_A$ to $\approx 300.9$, while $r_{s,drag}$ remains comparatively loosely determined, reaching $144^{+11}_{-8.7}$ Mpc for the full combination. Because in this analysis we employ only the acoustic-scale prior $\ell_A$ rather than the full set of CMB distance priors, the early-universe physics is less tightly anchored: a broader range of $r_{s,drag}$ remains open, and the suppression of the interaction---although still effective---is weaker than that obtained with the complete CMB prior set.

\textit{\label{Con}Conclusion} --- In this work, we have explored a new avenue within dark sector physics to address the open questions of the standard $\Lambda$CDM model. Unlike the ``top-down'' phenomenological approaches prevalent in the literature---where the interaction term is often assumed to be proportional to the Hubble parameter ($Q \propto H\rho$) based merely on dimensional analysis---we have proposed a ``bottom-up'' model grounded in microphysics. By drawing an analogy with the Boltzmann collision term, we formulated an interaction $Q = -A\frac{H_0}{\bar\rho_{\text{c},0}}\bar{\rho}_{\text{DM}}^2+B\frac{H_0}{\bar\rho_{\text{c},0}}\bar{\rho}_{\text{DE}}^2$ that describes a reversible process of particle annihilation and creation. This formulation provides a clear physical interpretation for the  {effective} coupling coefficients as  {effective} thermally-averaged cross-sections per unit mass and avoids the theoretical issues of non-locality associated with $H$-dependent interactions.

The constraints on the Hubble constant obtained from the different data combinations are mutually consistent within the framework of our model, with central values ranging from $\sim 67$ km s$^{-1}$ Mpc$^{-1}$ for the late-time combinations to $\sim 70$ km s$^{-1}$ Mpc$^{-1}$ when BAO is included. The combined analysis yields a measurement of $H_0 = 68.9^{+1.5}_{-2.0}$ km s$^{-1}$ Mpc$^{-1}$. A crucial finding of our analysis is the stringent constraint on the interaction strength. Contrary to models that require significant coupling to fit the data, we found strict upper limits for both interaction coefficients at the 95\% confidence level. For the combined dataset, the interaction coefficient $A$ is constrained to $A < 8.04\times10^{-21}$, while the coefficient $B$ is limited to $B < 4.28\times10^{-3}$ at 95\% CL. Crucially, the combination of these limits with the observational bound on $H_0$ translates directly into an  {upper limit} on the thermally-averaged annihilation cross-section per unit of mass.  Due to the cosmic expansion rate and the vast characteristic size of the Universe, the measured cross-section today appears as a small, yet non-null value, which stimulates the search for more precise limits using complementary data sources.

We also performed a  {goodness-of-fit analysis} using the reduced chi-squared statistic. We found that the IDE model yields $\chi^2_{\nu} \approx 0.88$, a value highly comparable to that of the standard $\Lambda$CDM model ($0.8793$). This demonstrates that the proposed IDE scenario is a viable cosmological model that fits the observational data with high precision.

Finally, we observed a correlation between the interaction strength and the comoving sound horizon at the drag epoch, $r_{s,drag}$. While the background probes allow for larger deviations in $r_{s,drag}$, the inclusion of the acoustic-scale prior $\ell_A$ partially breaks this degeneracy, constraining the sound horizon to $\approx 144$ Mpc and suppressing the allowed interaction range. Future perspectives for this work include extending the analysis to linear perturbations to investigate if the microphysical nature of this coupling affects the growth of structure ($f\sigma_8$) and the full CMB power spectrum distinctively from standard $\Lambda$CDM.

\textit{Acknowledgments} --- AAE and FBA acknowledge the support from the University of Science and Technology of China. FBA also acknowledges support from the Chinese Academy of Sciences as well as the Br-A Talent Program.

\appendix

\section{Microphysical Foundations of the
Dark-Sector Interaction Term}
\label{app:micro}

This appendix establishes the validity and the physical context of the
elementary $2\!\to\!2$ process
\begin{equation*}
\chi + \chi \;\longleftrightarrow\; \phi + \phi ,
\end{equation*}
that underlies the quadratic interaction term of Eq.~\eqref{Q1}. We show
that this process originates from a renormalizable Lagrangian containing a
heavy scalar mediator of mass $M$, written separately for the scalar and
vector realizations of $\chi$ and $\phi$, compute the resulting
thermally-averaged cross section, and make explicit the conditions under
which the microscopic reaction maps onto the coarse-grained fluid source
$Q\propto\bar\rho^2$ that enters the continuity
equations~\eqref{eqDM}--\eqref{eqDE} of the main text. The
$2\!\leftrightarrow\!2$ reaction is the natural starting point: among the
local number-changing processes $n\,\chi\leftrightarrow m\,\phi$,
reversibility excludes the single-particle decays $\chi\!\to\!\phi$ (which
are not collisional and yield a $Q\propto\bar\rho$ term), while higher
multiplicities such as $2\!\leftrightarrow\!3$ or $3\!\leftrightarrow\!3$
correspond to operators of higher mass dimension, are phase-space
suppressed, and are subleading at weak coupling, so that
$\chi+\chi\leftrightarrow\phi+\phi$ is the simplest collisional,
reversible, number-changing reaction compatible with a local and
renormalizable description.

To realise this scenario concretely we impose a
$\mathbb{Z}_2\times\mathbb{Z}_2$ symmetry, $\chi\!\to\!-\chi$ and
$\phi\!\to\!-\phi$, acting independently on the two dark fields: it
stabilises each species against single-particle decay (forbidding the
linear channels that would generate a $Q\propto\bar\rho$ term) and forbids
every odd operator, so that the lowest-dimensional Lorentz-invariant
interaction is the quartic $\chi^2\phi^2$. The reaction
$\chi+\chi\leftrightarrow\phi+\phi$
is therefore the \emph{unique} leading number-changing process, converting
particles in pairs $2\chi\leftrightarrow2\phi$, and is exactly the
field-level origin of the quadratic source $Q\propto\bar\rho^2$.

The forward channel $\chi\chi\!\to\!\phi\phi$ is open only above
threshold, $\sqrt{s}\ge 2m_\phi$, and the inverse channel above
$\sqrt{s}\ge 2m_\chi$. In the cold regime relevant to the background
evolution the dark-matter quanta are non-relativistic, the centre-of-mass
energy sits just above threshold, and the partial-wave expansion is
dominated by the $s$-wave, so the amplitude is momentum independent,
$|\mathcal{M}|^2\simeq{\rm const}$, which is precisely what makes the
thermally averaged cross-section a single number
$\langle\sigma|v|\rangle_{A,B}$ and reduces the collision term to the
simple $\bar\rho^2$ form. The elementary reaction furnishes the
fluid-level source $Q$ only when the following conditions hold
simultaneously:
\begin{enumerate}
\item[(V1)] \emph{Kinetic equilibrium within each species.} The
self-interactions of $\chi$ (and of $\phi$) are fast compared with the
inter-species conversion rate, so that each phase-space distribution
tracks its own number density and the pair-conversion rate is governed by
a single thermally-averaged cross section $\langle\sigma|v|\rangle_{A,B}$.
This is what fixes the quadratic $\bar\rho^2$ dependence of the source.
 {As shown below in Eq.~\eqref{eq:Leff}, the \emph{same} mediator
that generates the conversion process also generates the $\chi^4$ and
$\phi^4$ self-couplings able to sustain this equilibrium, so the operators
are present by construction. 
Their existence is thus a necessary structural ingredient rather than an
automatic guarantee.}
\item[(V2)] \emph{Non-relativistic dark matter,} $T_\chi\ll m_\chi$,
so that $\bar\rho_\chi=m_\chi n_\chi$ and $P_\chi\simeq0$. This is the
single approximation that allows the source $Q$ to be expressed in terms
of the mass densities through $\bar\rho=m\,n$, with relative error
$\mathcal{O}(T_\chi/m_\chi)$.
\item[(V3)] \emph{A particle-like dark-energy excitation.} The
field $\phi$ must admit localised quanta able to scatter; a strict
cosmological constant is excluded. This is implemented by the decomposition
$\bar\rho_{\rm DE}=\rho_{\rm vac}+m_\phi n_\phi$, in which $Q$ is sourced
only by the sub-dominant particle-like component while $w\simeq-1$ is
driven by $\rho_{\rm vac}$.  {We denote its fraction by
$f_{\rm part}\equiv\bar\rho_{\phi,{\rm part}}/\bar\rho_{\rm DE}=m_\phi
n_\phi/\bar\rho_{\rm DE}$, so that the inverse channel sources
$n_\phi^2=(f_{\rm part}\bar\rho_{\rm DE}/m_\phi)^2$ and the coefficient
$B$ multiplying $\bar\rho_{\rm DE}^2$ is effective, carrying a factor
$f_{\rm part}^2$.}
\item[(V4)] \emph{Weak coupling,} $A,B\ll1$, confirmed \emph{a
posteriori} by the bounds reported in the main text.  {Assuming perturbative microscopic couplings,} this justifies the
use of the tree-level amplitude in the cross section computed below.
\end{enumerate}
Under conditions (V1)--(V4) the microscopic reaction maps onto
the coarse-grained interaction $Q\propto\bar\rho^2$; outside this regime
(relativistic $\chi$, strong coupling, or a non-dynamical $\phi$) the
mapping breaks down and a different functional form of $Q$ is required.
To exhibit a renormalizable origin for the quartic operator, we introduce
a real scalar mediator $S$ of mass $M$ coupling to the two
$\mathbb{Z}_2\times\mathbb{Z}_2$-even bilinears:
\begin{align}
\mathcal{L} &= \mathcal{L}_\chi+\mathcal{L}_\phi
   +\mathcal{L}_S+\mathcal{L}_{\rm int} ,\label{eq:Ltot}\\
\mathcal{L}_\chi &= \tfrac12(\partial_\mu\chi)(\partial^\mu\chi)
   -\tfrac12 m_\chi^2\chi^2 ,\\
\mathcal{L}_\phi &= \tfrac12(\partial_\mu\phi)(\partial^\mu\phi)-V(\phi),\\
\mathcal{L}_S &= \tfrac12(\partial_\mu S)(\partial^\mu S)
   -\tfrac12 M^2 S^2 ,\\
\mathcal{L}_{\rm int} &= -\tfrac12\, g_\chi\, S\,\chi^2
         -\tfrac12\, g_\phi\, S\,\phi^2 .
\label{eq:Lint_med}
\end{align}
The trilinear couplings $g_\chi,g_\phi$ have mass dimension $+1$, so
each operator is dimension-four and the theory is renormalizable. Note
that $S$ couples to $\chi^2$ and to $\phi^2$ but \emph{not} to
$\chi\phi$: there is no $S\chi\phi$ vertex, a single fact that
controls the matching below.
The Euler--Lagrange equation for $S$ is
\begin{equation}
(\Box+M^2)\,S = -\tfrac12 g_\chi\chi^2-\tfrac12 g_\phi\phi^2
\equiv -J .
\end{equation}
For processes with characteristic momenta well below the mediator mass,
$|\Box|\sim p^2\ll M^2$, the kinetic term is negligible and
\begin{equation}
S \;\simeq\; -\frac{J}{M^2}
  = -\frac{1}{2M^2}\bigl(g_\chi\chi^2+g_\phi\phi^2\bigr).
\label{eq:Ssol}
\end{equation}
Substituting~\eqref{eq:Ssol} back into~\eqref{eq:Ltot} (the standard
tree-level integrating-out, exact because $J$ is independent of $S$)
gives the effective Lagrangian
\begin{align}
\mathcal{L}_{\rm eff}
&= \frac{J^2}{2M^2}
 = \frac{1}{8M^2}\bigl(g_\chi\chi^2+g_\phi\phi^2\bigr)^2 \nonumber\\
&= \underbrace{\frac{g_\chi^2}{8M^2}\chi^4}_{\text{DM self}}
 + \underbrace{\frac{g_\phi^2}{8M^2}\phi^4}_{\text{DE self}}
 + \underbrace{\frac{g_\chi g_\phi}{4M^2}\,\chi^2\phi^2}_{\text{cross}} .
\label{eq:Leff}
\end{align}
The cross term is the operator we want. Matching it to the contact
form $-\tfrac14\lambda_{\rm eff}\chi^2\phi^2$ gives
\begin{equation}
{\;\mathcal{L}_{\rm int}^{\rm eff}=-\frac{\lambda_{\rm eff}}{4}\,\chi^2\phi^2,
\qquad
\lambda_{\rm eff}= \frac{g_\chi\, g_\phi}{M^2}\;}
\label{eq:lambda_eff}
\end{equation}
(the sign is fixed by the relative sign of $g_\chi g_\phi$; we take
$\lambda_{\rm eff}>0$). The quartic coupling $\lambda$ of the contact
theory is thus \emph{not} fundamental: it is the ratio of a coupling
product to the squared mediator mass, $\lambda_{\rm eff}\propto M^{-2}$.

Equation~\eqref{eq:Leff} simultaneously produces the quartic
self-couplings $\lambda_\chi=g_\chi^2/(2M^2)$ and
$\lambda_\phi=g_\phi^2/(2M^2)$. These are exactly the
intra-species interactions that can maintain the kinetic equilibrium
required by~(V1).  {The mediator that drives the conversion therefore
also supplies the self-scattering operators; whether these keep each
species in kinetic equilibrium is a dynamical question, requiring the
corresponding rates to exceed both $H$ and the conversion rate, rather
than a property guaranteed by the existence of the operators alone.}
So far we treated $\chi$ and $\phi$ as real scalars. Each dark species
may equally well be a massive vector, and it is instructive to write the
two realisations separately and then show that they lead to the
\emph{same} fluid-level interaction. In the purely scalar realisation the
mediator couples to the scalar bilinears,
\begin{align}
\mathcal{L}_{\rm s} =\;
&\underbrace{\tfrac12(\partial\chi)^2-\tfrac12 m_\chi^2\chi^2}
   _{\text{scalar DM}}
+\underbrace{\tfrac12(\partial\phi)^2-V(\phi)}_{\text{scalar DE}}
\nonumber\\[2pt]
+&\underbrace{\tfrac12(\partial S)^2-\tfrac12 M^2 S^2}_{\text{mediator}}
\;-\;\tfrac12 S\big(g_{\chi s}\,\chi^2+g_{\phi s}\,\phi^2\big),
\label{eq:Lscalar}
\end{align}
whereas in the purely vector realisation, with massive vector fields
$X_\mu$ (dark matter) and $Y_\mu$ (dark energy) and field strengths
$X_{\mu\nu}=\partial_\mu X_\nu-\partial_\nu X_\mu$,
$Y_{\mu\nu}=\partial_\mu Y_\nu-\partial_\nu Y_\mu$, it couples to the
vector bilinears,
\begin{align}
\mathcal{L}_{\rm v} =\;
&\underbrace{-\tfrac14 X_{\mu\nu}X^{\mu\nu}
   +\tfrac12 m_X^2 X_\mu X^\mu}_{\text{vector DM}}
+\underbrace{-\tfrac14 Y_{\mu\nu}Y^{\mu\nu}
   +\tfrac12 m_Y^2 Y_\mu Y^\mu}_{\text{vector DE}}
\nonumber\\[2pt]
+&\underbrace{\tfrac12(\partial S)^2-\tfrac12 M^2 S^2}_{\text{mediator}}
\;-\;\tfrac12 S\big(g_{\chi v}\,X_\mu X^\mu+g_{\phi v}\,Y_\nu Y^\nu\big).
\label{eq:Lvector}
\end{align}
In both cases the mediator couples to a $\mathbb{Z}_2\times\mathbb{Z}_2$
even bilinear of dark matter and a bilinear of dark energy, with no
direct $\chi$--$\phi$ ($X$--$Y$) vertex. Integrating out the heavy
$S$ as before, $S\simeq-(2M^2)^{-1}(g_{\chi}\mathcal{O}_\chi
+g_{\phi}\mathcal{O}_\phi)$, yields in each case an effective
cross operator of identical structure,
\begin{equation}
\footnotesize
\mathcal{L}_{\rm eff}^{\rm cross}
=\frac{g_\chi\,g_\phi}{4M^2}\,\mathcal{O}_\chi\,\mathcal{O}_\phi ,
\quad
\begin{cases}
\mathcal{O}_\chi=\chi^2,\;\mathcal{O}_\phi=\phi^2 & \text{(scalar)}\\[1pt]
\mathcal{O}_\chi=X_\mu X^\mu,\;\mathcal{O}_\phi=Y_\nu Y^\nu & \text{(vector)}
\end{cases}
\label{eq:Leff_both}
\end{equation}
so that the conversion is governed by a single effective constant
$\lambda_{\rm eff}=g_\chi g_\phi/M^2$ in either realisation. The two
descriptions are therefore equivalent at the level relevant here: each
mediates a reversible $2\!\leftrightarrow\!2$ process
($\chi\chi\leftrightarrow\phi\phi$ or $XX\leftrightarrow YY$) and produces
a fluid-level source of the same quadratic form $Q\propto\bar\rho^2$. The
only differences are the polarisation-averaging factors  {(from the sums over physical polarisations)}
and the mass dictionary $n\to\bar\rho$ appropriate to each bilinear. At
the background level both collapse onto the single phenomenological pair
$(A,B)$ of the main text, which is thus an effective coarse-graining over
whichever spin assignment is realised in the dark sector. A Majorana
fermionic dark-matter component would enter through the operator
 {$S\bar\chi\chi$}; the mediator couples to the dimension-four
bilinear $S\,\bar\chi\chi$, but  {after integrating out $S$, the scale $\Lambda_c$ suppresses the non-renormalizable effective operator $(\bar\chi\chi)\phi^2/\Lambda_c$}. The same macroscopic chain
nonetheless follows, so we restrict the explicit treatment to the
scalar and vector cases above.

Because $S$ has no $S\chi\phi$ vertex, the only tree diagram for
$\chi\chi\!\to\!\phi\phi$ is the $s$-channel exchange
$\chi\chi\!\to\!S\!\to\!\phi\phi$ (there are no $t$- or $u$-channel
graphs):
\begin{equation}
\mathcal{M} = -\,\frac{g_\chi g_\phi}{s-M^2}
\;\xrightarrow[\;M^2\gg s\;]{}\;
\frac{g_\chi g_\phi}{M^2}=\lambda_{\rm eff},
\end{equation}
so $|\mathcal{M}|^2\simeq\lambda_{\rm eff}^2$, reproducing the
contact-theory result $|\mathcal{M}|^2=\lambda^2$ with
$\lambda\to\lambda_{\rm eff}$. Near threshold, $s=4m_\chi^2$, the
heavy-mediator (contact) description requires $M\gg2m_\chi$; for
$M\sim2m_\chi$ the $s$-channel resonates and the contact approximation
fails. Because the two  {final-state $\phi$ particles} are identical, the
$s$-wave thermally averaged cross-section carries the corresponding
$1/2$ symmetry factor relative to the distinguishable-particle case.

 {Retaining the final-state two-body phase space, it reads}

\begin{equation*}
\langle\sigma|v|\rangle_A\simeq\frac{\lambda_{\rm eff}^2}{64\pi m_\chi^2}\,
\sqrt{1-\frac{m_\phi^2}{m_\chi^2}}\,,
\end{equation*}
 {the square root being the final-state phase-space factor, real
only for $m_\chi\ge m_\phi$. In the hierarchical regime $m_\phi\ll m_\chi$
adopted throughout---consistent with a light dark-energy
excitation---this factor reduces to unity, giving}
\begin{equation}
\langle\sigma|v|\rangle_A\simeq\frac{\lambda_{\rm eff}^2}{64\pi m_\chi^2}
=\frac{(g_\chi g_\phi)^2}{64\pi\, m_\chi^2\, M^4}.
\label{eq:svA_med}
\end{equation}
 {The inverse channel analogously gives
$\langle\sigma|v|\rangle_B\simeq[\lambda_{\rm eff}^2/(64\pi m_\phi^2)]
\sqrt{1-m_\chi^2/m_\phi^2}$, which is open only for $m_\phi\ge m_\chi$;
the two channels are therefore not simultaneously unsuppressed in the
strict cold limit unless $m_\chi\simeq m_\phi$, a point we return to when
interpreting $B$ below.}

\subsection{Consequence for the observable coefficients}

Equation~\eqref{eq:svA_med} completes the microphysical input. Inserting
it into the background interaction term [Eq.~\eqref{Q1}] through the
non-relativistic identification $\bar\rho=m\,n$ fixes the dictionary
between the dimensionless coefficients and the underlying cross sections,
$A=\langle\sigma|v|\rangle_A\,\bar\rho_{c,0}/(m_\chi H_0)$ and
 {$B=f_{\rm part}^2\,\langle\sigma|v|\rangle_B\,m_\chi\bar\rho_{c,0}/(m_\phi^2 H_0)$,
the factor $f_{\rm part}^2$ reflecting that only the particle-like fraction
of dark energy participates in the inverse reaction [cf.\ (V3)]}. Using
the mediator result $\lambda\to\lambda_{\rm eff}=g_\chi g_\phi/M^2$ of
Eq.~\eqref{eq:lambda_eff}, this gives
\begin{equation}
A\simeq\frac{\bar\rho_{c,0}\,(g_\chi g_\phi)^2}{64\pi\,H_0\,m_\chi^3\,M^4},
\qquad
B\simeq \bm{f_{\rm part}^{2}}\,\frac{\bar\rho_{c,0}\,(g_\chi g_\phi)^2\,m_\chi}{64\pi\,H_0\,m_\phi^4\,M^4}.
\label{eq:AB_med}
\end{equation}
 {Consequently a bound on $B$ constrains the product
$f_{\rm part}^2\,\langle\sigma|v|\rangle_B$ rather than the inverse cross
section alone, and cannot be converted into a microscopic
$\langle\sigma|v|\rangle_B$ without specifying the corpuscular fraction
$f_{\rm part}$ of the dark-energy sector.}
The key structural result is $A,B\propto M^{-4}$. The observed
smallness of the coupling is then \emph{explained}, not assumed: the
$95\%$ upper limit $A<8.04\times10^{-21}$ translates into a \emph{lower
bound} on the mediator mass,
\begin{equation}
M \;\gtrsim\;
\left[\frac{\bar\rho_{c,0}\,(g_\chi g_\phi)^2}{64\pi\,H_0\,m_\chi^3\,A_{\max}}
\right]^{1/4},
\label{eq:Mbound}
\end{equation}

i.e.\  {for fixed couplings and masses and within the contact regime
$M\gg2m_\chi$, a tiny $A$ translates into a \emph{lower bound} on the
mediator mass}, in direct analogy with the suppression $G_F\sim g^2/M_W^2$
of the weak interaction.  {The bound constrains the combination
$(g_\chi g_\phi)^2/(M^4 m_\chi^3)$ rather than $M$ alone; under those
assumptions it} converts the phenomenological bound on $A$ into a
statement about the scale of new physics in the dark sector, derived
from a well-defined Lagrangian rather than being inserted by hand.

\section{Details of the Statistical Analysis\label{API}}

We calculated the reduced chi-squared value, $\chi^2_{\nu}$, for each data combination analyzed, for both the proposed IDE model and the standard F$\Lambda$CDM model. The values are listed in Table \ref{tab:chired}. For the full combination of datasets (PP $+$ $\ell_A$ $+$ BAO $+$ CCs), the IDE model yields $\chi^2_{\nu}=0.8756$, marginally below the value obtained for the $\Lambda$CDM model ($0.8793$); the two models therefore provide fits of essentially identical quality when all datasets are combined. The same near-degeneracy is observed for the late-time combinations PP $+$ CCs, PP $+$ $\ell_A$ $+$ CCs, and PP $+$ BAO $+$ CCs, for which the IDE and $\Lambda$CDM values differ only at the third decimal place and remain well below unity, indicating a good fit to these observations. The only combination for which the IDE model fits appreciably worse than $\Lambda$CDM is BAO $+$ $\ell_A$, with $\chi^2_{\nu}=1.1247$ against $0.9349$; this mild residual  {indicates a weaker fit for} the background-only BAO $+$ $\ell_A$ combination to accommodate the interaction, and it is washed out once the supernova and cosmic-chronometer data are included, the full combination recovering a fit fully comparable to $\Lambda$CDM.

\begin{table}[H]
    \centering
    \setlength{\tabcolsep}{6.0pt}
    \caption{$\chi^2_{\nu}$ of each data combination for the IDE model and for the F$\Lambda$CDM model.}
    \begin{tabular}{l| c c}
    \hline\hline
       Data  &  $\chi^2_{\nu, \text{IDE}}$& $\chi^2_{\nu,\text{F$\Lambda$CDM}}$ \\
       \hline
         PP + CCs & 0.8775 & 0.8761 \\
         BAO + $\ell_A$ & 1.1247 & 0.9349\\
         PP + $\ell_A$ + CCs & 0.8770& 0.8748\\
         PP + BAO + CCs & 0.8761& 0.8773\\
         PP + $\ell_A$ + BAO + CCs & 0.8756 &0.8793\\\hline 
    \end{tabular}
    \label{tab:chired}
\end{table}

\bibliography{apssamp}

\end{document}